# Tachyons or Antiparticles?

J.N. Pecina-Cruz

Department of Physics
The University of Texas-Pan American
1201 West University Drive
Edinburg, Texas 78541-2999
jpecina2@panam.edu

This article proves that elementary particles can violate the second postulate of the Special Theory of Relativity. According to Heisenberg's principle, superluminal particles are allowed in the theoretical framework of Quantum Mechanics. In this paper the tachyons, are identified with the unitary irreducible representations of the "full" Poincare group (Poincare group plus space-time reflections.) on the past time-like cone.

**Introduction**

Einstein's theory of relativity demands that exist a symmetry that leaves invariant the laws of physics [1], under the relative motion of reference frames. Einstein also claims that the time coordinate rests on the same footing than the space coordinates. Therefore the Poincare group must be enlarged to include space-time inversions. Ne'eman et al [2] have conjectured that a theory of quantum gravity could be the result of enhancing the Poincare group to SA(r,4). Schwinger already intended to introduce the strong reflections in quantum field theory [3]. Section 1 is devoted to explaining the tunneling of a particle from the future time-like region to the past time-like region of the space-time. In sections 2 to 6 the unitary irreducible representations of the "full" Poincare group are constructed.

**1. The Interpretation of Tachyons as Antiparticles**

In the one particle scheme Feynman and Stückelberg interpret antiparticles as particles moving backwards in time [4]. This argument is reinforced by G. Sudarshan [5] and S. Weinberg [6] who realized that antiparticles existence is a consequence of the violation of the principle of causality in quantum mechanics. The temporal order of the events could be inverted when a particle wanders in the neighborhood of the light cone. How is the antimatter generated from matter? According to Heisenberg's uncertainty principle a particle wandering in the neighborhood of the light-cone suddenly tunnels (acquire a speed greater than the speed of light) from the time-like region to the space-like region; in this region the relation of cause and effect collapses. Since if an event, at $x_2$ is observed by an observer A, to occur later than one at $x_1$, in other words $x_2^0 > x_1^0$. An observer B moving with a velocity **v** respect to observer A, will see the events separated by a time interval given by

$$x_2'^0 - x_1'^0 = L^0_\alpha(v)(x_2^\alpha - x_1^\alpha), \tag{1}$$

Where $L^\beta_\alpha(v)$ is a Lorentz boost. from equation (1), it is found that if the order of the events is exchanged for the observer B, that is, $x_2'^0 < x_1'^0$ (the event at $x_1$ is observed later than the event at $x_2$.), then a particle that is emitted at $x_1$ and absorbed at $x_2$ as observed by A, it is observed by B as if it were absorbed at $x_2$ before the particle were emitted at $x_1$. The temporal order of the particle is inverted. This event is completely feasible in the neighborhood of the light-cone, since the uncertainty principle allows a particle tunnel from time-like to space-like cone regions. That is the uncertainty principle will consent to the space-like region reach values above than zero as is shown by equation:

$$(x_1 - x_2)^2 - (x_1^0 - x_2^0)^2 \leq \left(\frac{\hbar}{mc}\right)^2,$$

Where $\left(\frac{\hbar}{mc}\right)^2 > 0,$ (2)

And $\frac{\hbar}{mc}$ is the Compton wave-length of the particle. The left hand side of the equation (2) can be positive or space-like for distances less or equal than the square of the Compton wavelength of the particle. Therefore, causality is violated. The only way of interpreting this phenomenon is assuming that the particle absorbed at $x_2$, before it is emitted at $x_1$ as it is observed by B, is actually a particle with negative mass and energy, and certain charge, and spin, moving backward in time; that is $t_2 < t_1$ [6]. This event is equivalent to see an antiparticle moving forward in time with positive mass and energy, and opposite charge and spin that it is emitted at $x_1$ and it is absorbed at $x_2$. With this reinterpretation the causality is recovered. But, this interpretation is still incomplete, since these particles do not have a real existence. Imaginary masses, length contraction, and time dilations populate the theory of tachyons [5]. This paper interprets the tachyons as particles that tunnel from the future time-like to the past time-like regions. So, superluminal particles are nothing else, but certainly antiparticles. Heisenberg's uncertainly principle limits their existence to very short life-time.

The square of the mass for the observers A and B is a Poincare invariant. In other words, *the mass of a particle itself is not an invariant* [7], *but its square* $p^2 = -m^2$. There is nothing to prevent that the rest mass could have different sign in another reference frame, as happen with its energy, electric charge and spin. In recent publication [8], it is conjectured that a particle moving backward in time possesses a negative mass. When this particle is observed as an antiparticle moving forward in time its mass is positive. Still its mass square is a Poincare invariant.

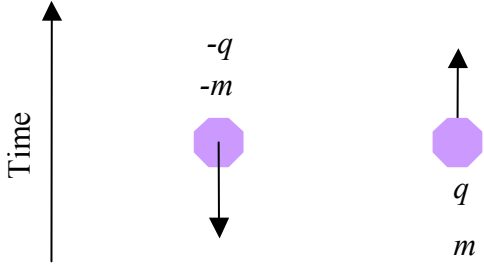

Figure 1. A particle moving backwards in time is equivalent to an antiparticle moving forward in time with opposite spin (not showed), charge, and mass.

Under the light of the uncertainty principle the arguments against the negative energy states are untenable. Since a particle must reach a speed greater than that of the light to reverse the time direction and acquire a negative mass. Classically this event is impossible, but in quantum mechanics it is totally feasible. Therefore, the potential barrier that has to overcome a particle to be seen as an antiparticle is closely related with the impossibility to reach the speed of light. Let us picture this phenomenon with Klein's example. Assume a potential of the form

Region I: $E - V = c\sqrt{|\vec{p}|^2 + m_0^2 c^2} > m_0 c^2$ (3)

Region II: $m_0 c^2 > E - V = c\sqrt{|\vec{p}|^2 + m_0^2 c^2} > -m_0 c^2$

Region III: $-m_0 c^2 > E - V = c\sqrt{|\vec{p}|^2 + m_0^2 c^2}$,

Klein's argument is useful to visualize the violation of causality on the neighborhood of the light-cone, and to understand the creation of antiparticles when those particles move space-time distances on the order of their Compton wavelength near to the light-cone boundaries [9]. According to Klein's paradox the index of reflected particles is greater than one. The paradox disappears if one takes into account the creation of particles and antiparticles. The particles move in the direction of reflection and the antiparticles are attracted in opposite direction by the potential V. According to equation (2), tunneling can occur from region I to III.

In Region II the particle possess an imaginary mass, $p_\mu p^\mu = E^2 - |\vec{p}|^2 = M^2 < 0$, and hence an imaginary momentum. During this time, the particle is "off-mass-shell," the particle exists as a virtual particle and it is responsible for force transmissions. When finally the particle reaches Region III, passes from a virtual to a real existence, with its mass and energy negatives [9] [10]. To pass from Region I to Region II or III is not a "downhill" transition, since the particle has to reach a speed greater than the speed of light, when moves from Region I to Region II, that is, it has to undergo an infinite acceleration. But an infinite acceleration is classically impossible. However, Heisenberg's uncertainty principle makes this event feasible.

In this paper the irreducible representations of the Poincare group are reviewed to include simultaneous time-space reflections. The irreducible representations of the Poincare group with simultaneous reflection of time and space will be constructed.

It is also conjectured in this paper that all the regions formed by the light cone have a physical meaning and that these regions describe the elementary particles. Lamb's shift is an example of the physical existence of one of those regions [11].

The possible existence of dark matter could be a consequence of the negative mass value of the four-dimensional mirror particles reflection. Massless particles have an infinite Compton wavelength, therefore their particles and antiparticles appear in the same proportion. Photons and antiphotons would appear in the same amount. Neutrinos would appear in nature in asymmetric proportion, since these are not massless particles. At distances compared to the Compton wavelength of system of particles the value of the entropy of a close system could decrease due to the arrow of time that can be reversed.

## 2. Unitary Irreducible Representations of Poincare Group with Simultaneous Space-Time Reflections

The unitary irreducible representations of the extended group of Poincare have been constructed by constraining from the onset the energy values to be positive as a consequence of this assumption time reversal operator is antilinear. Later the antilinear property of the time reversal operator is utilized to prevent the existence of negative energy states for elementary particles [12]. In this paper, however the opposite approximation is proposed. The quantum negative energy states are accepted on the basis that the creation of an antiparticle (particle of negative energy) is briefly allowed by an interval of time ruled by the uncertainty principle; while the particle moves around the light cone boundaries that separate time-like from space-like regions. On the neighborhood of this boundary the probability of tunneling is greater than zero. A collapse of particles toward negative energy states is classically prohibited by the impossibility that a particle can exceed the speed of light. However, in quantum mechanics is briefly permitted by the Heisenberg uncertainty principle [see eqn. (2)]; as it is discussed above. If this phenomenon occurred continuously, the propagation speed of the information between two objects would be infinite. Furthermore, tunneling has a low probability of occurrence, and the mean lifetime of an antiparticle is in general short, since this is immediately absorbed by its particle to transform into gamma rays.

The classical way to construct the unitary irreducible representations of the Poincare group is by using the technique of the little group. Let us consider the orbits or regions where the magnitude of the four vector $\hat{s}^2 = x^2+y^2+z^2-c^2t^2$ is: zero, with all its components equal to zero; zero with its components different of zero; greater than zero, and less than zero. To obtain these unitary irreducible representations we will use the two dimensions

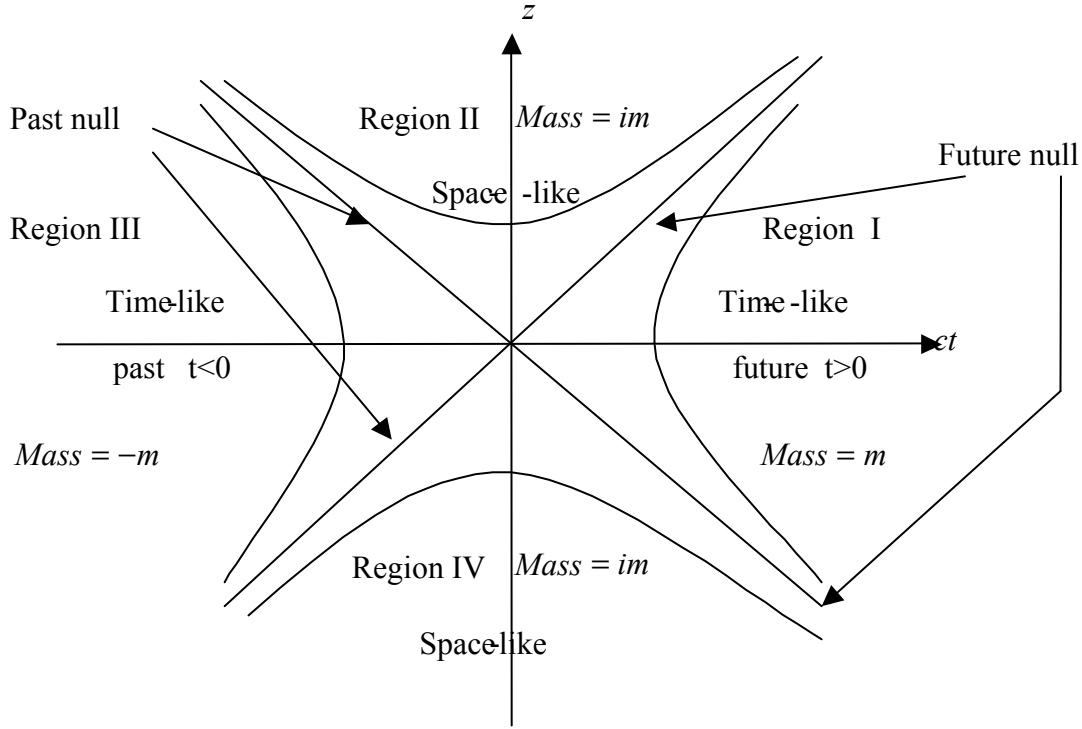

Figure 2

space-time t-z of Figure 2, in the momentum representation. Therefore the six regions to be discussed are

1. ŝ. ŝ < 0,                                                              space-like

2. ŝ. ŝ > 0, t > 0,                                                      future time-like

3. ŝ. ŝ > 0, t < 0,                                                      past time-like

4. ŝ. ŝ = 0, ŝ ≠ 0, t > 0,                                             future null

5. ŝ. ŝ = 0, ŝ ≠ 0, t < 0,                                             past null

6. ŝ = 0,                                                                  the zero vector

The representation for the Poincare group with simultaneous space-time inversions will be constructed following references [13].

Let us choose the vectors $|\hat{k}\rangle$ as the basis vectors of an irreducible representation $T^{(k)}$, of the subgroup of translations of the Poincare group. Then the basis vectors $|\hat{k}'\rangle$, with $\hat{k}' = W\hat{k}$, and $W$ a Lorentz transformation, lie in the same representation. That is, $k'$ and $k$ are in the same region of the space-time; these vectors have the same "length", (k',k') = (k,k).

$$T(\hat{u})T(W)|\hat{k}\rangle = T(W)|\hat{k}\rangle \exp(W\hat{k},\hat{u}) \qquad (4)$$

The vectors $T(W)|\hat{k}\rangle$ transform according to the irreducible representation $T^{(Wk)}$, of the subgroup of translations.

Let us start our analysis with the future time-like cone or region 2. By choosing the time-like four-vector $\hat{k}_0 = (0,0,0,k)$ with $k>0$, it is found that the little group, corresponding to the orbit of the point $(0,0,0,k)$, is the group of rotations in three dimensions SU(2). Therefore, the representation is labeled by $\hat{k}_0$ and by an irreducible representation label of the three dimensional rotational group SU(2). Starting from the 2s + 1 basis vectors of the little group SU(2) the bases vectors $|\hat{k}sm_s\rangle$, of an irreducible representation of the Poincare group, are generated by a pure Lorentz transformation (a boost) which carries $\hat{k}_0$ into $\hat{k}$, that is $W_k\hat{k}_0 = \hat{k}$. The basis vectors of the representation are given by

$$|\hat{k}sm_s\rangle = T(W_k)|\hat{k}_0 sm_s\rangle. \qquad (5)$$

This group operation preserves the "length" of the vector $\hat{k}_0$, that is $(k,k) = (k_0,k_0)$. By applying a Lorentz transformation followed by a translation to equation (55) can be proved that the vectors $|\hat{k}sm_s\rangle$ furnish a unitary irreducible representation of the Poincare group

$$T(\hat{u})T(W)|\hat{k}sm_s\rangle = |\hat{k}'sm'_s\rangle D_{m'_s m_s}(W_{k_0}) \exp(\hat{k}',\hat{u}). \qquad (6)$$

Where $\hat{k}' = W\hat{k}$, and $(W\hat{k}, W\hat{k}) = (\hat{k}, \hat{k})$. $D_{m'_s m}(W_{k_0})$ are the unitary irreducible representations of the little group SU(2), $W$ is an arbitrary Lorentz transformation. The representations are unitary because the generators of the group are unitary, and irreducible because the bases vectors of the representations are generated from a single vector $|\hat{k}_0 sm_s\rangle$, with linear momentum equal to zero. Their equivalent unitary irreducible representations are denoted by $P^{(k_0,s)}$. These representations have the same length of the four-vector $\hat{k}$ than that of $\hat{k}_0$. The set of inequivalent representations with different magnitude of the four-vector $\hat{k}$ are denoted by $P^{(k,s)}$.

Let us construct the unitary irreducible representations of the Poincare group with simultaneous space-time reflections. First note that if the vectors $|\hat{k}\rangle$ are the bases of an irreducible representation of the translation group, and $I$ is the space-time reflection operator, then

$$T(\hat{u})T(I)|\hat{k}\rangle = T(I)T(I\hat{u})|\hat{k}\rangle = T(I)|\hat{k}\rangle\exp(I\hat{k},\hat{u}) \qquad (7)$$

Therefore, the vectors $T(I)|\hat{k}\rangle$ transform according to the $T^{(Ik)}$ representation of the translation subgroup of the Poincare group. But, the operator $I$ reverses, the space and time components of $\hat{k}$ that is, the energy $E = \hbar c k_t$ becomes negative. On the rest frame of the $P^{(k,s)}$ representation, the rest mass, $m_0 = \dfrac{\hbar I k_0}{c} = -\dfrac{\hbar k_0}{c}$, of such a particle would also be negative. There is nothing to prevent this from happening, since $m_0^2$ is a Poincare invariant [7], but not $m_0$. That is, if one observer sees a particle with rest mass positive another observer on an inertial frame could see the same particle with negative rest mass. An electron with negative rest mass and energy moving backwards in space and time could be interpreted as its antiparticle moving forward in space and time with positive rest mass and energy. This phenomenon could by explained by constructing the unitary irreducible representations of the Poincare group with simultaneous space-time inversions.

To construct the unitary irreducible representations of the full Poincare group one applies the space inversion followed by time inversion to the basis vectors $|\hat{k}sm_s\rangle$ of $P^{(k,s)}$. These basis vectors were generated from the special basis vector $|\hat{k}_0 sm_s\rangle$, where $\hat{k}_0 = (0,0,0,k)$. The space inversion $I_s$ leaves $\hat{k}_0$ invariant, and commutes with the generators of the little group SU(2). Therefore, the irreducible representations of the group, $Z_2 \times SU(2)$, can be labeled by the labels of the rotations, and reflection groups, namely the spin s, and the parity η. If one applies space inversions to the basis vector $|\hat{k}_0 sm_s \eta\rangle$ with definite parity label η = ±, one obtains

$$T(I_s)|\hat{k}_0 sm_s \eta\rangle = |\hat{k}_0 sm_s \eta\rangle \eta \qquad (8)$$

On the rest frame the parity, η, is an eigenvalue of the basis vector $|\hat{k}_0 sm_s \eta\rangle$.

Now, Let us start to construct the inequivalent unitary irreducible representations of the Poincare group with space inversions $P^{(k,s,\eta)}$. By applying these inversions to the basis vectors $|\hat{k}sm_s \eta\rangle$, given by equation (95), and including the parity label η one obtains

$$T(I_s)\left|\hat{k}sm_s\eta\right\rangle = T(I_s)T(W_k)\left|\hat{k}_0 sm_s\eta\right\rangle = T(W_{-k})T(I_s)\left|\hat{k}_0 sm_s\eta\right\rangle =$$
$$T(W_{-k})\left|\hat{k}_0 sm_s\eta\right\rangle\eta = \left|I_s\hat{k}sm_s\eta\right\rangle\eta \tag{9}$$

The Lorentz boost $W_{-k}$ changes the direction of space components of $\hat{k}$. Equation (99) shows that the basis vector $\left|\hat{k}sm_s\eta\right\rangle$ is an invariant space under space reflections; the action of the space inversion operator on a general basis vector leads to another basis vector. Hence, these bases vectors yield a representation for the Poincare group with space inversions.

If one considers the simultaneous action of the time-space inversions on a general basis vector one obtains

$$T(I)\left|\hat{k}sm_s\eta\right\rangle = T(I)T(W_k)\left|\hat{k}_0 sm_s\eta\right\rangle = T(W_k)T(I_t)T(I_s)\left|\hat{k}_0 sm_s\eta\right\rangle =$$
$$T(I_t)T(W_{-k})\left|\hat{k}_0 sm_s\eta\right\rangle\eta = T(I_t)\left|I_s\hat{k}'sm_s\eta\right\rangle\eta = \left|I\hat{k}'sm_s\eta\right\rangle\eta \tag{10}$$

The action of the full inversion operator on a general basis vector $\left|\hat{k}sm_s\eta\right\rangle$, which belongs to the unitary irreducible representation $P^{(k,s,\eta)+}$ generated from the vector $\hat{k}_0 = (0,0,0,k)$, leads to a basis vector in the $P^{(k,s,\eta)-}$ representation generated from the vector $I\hat{k}_0 = (0,0,0,-k)$. But, these two representations of the full Poincare group are equivalents. Therefore, the action of time inversion generates negative energy states, and also induces charge conjugation, as we will show it below.

### 3. Time Inversion and Charge Conjugation

If one applies linear time reflections to the basis vector $\left|\hat{k}_0 sm_s\eta\right\rangle$, one finds that

$$T(I_t)\left|\hat{k}_0 sm_s\eta\right\rangle = \left|I_t\hat{k}_0 sm_s\eta\right\rangle \tag{11}$$

Since, from equation (77 $T(I_t)\left|\hat{k}_0 sm_s\eta\right\rangle$ transforms according to the $T^{(I_t k_0 s)}$ representation of the subgroup of translations. That is, the basis vector $\left|\hat{k}_0 sm_s\eta\right\rangle$ with $\hat{k}_0 = (0,0,0,k)$ from the $P^{(k_0,s,\eta)+}$ representation is taken into the basis vector $\left|I_t\hat{k}_0 sm_s\eta\right\rangle$ with $I_t\hat{k}_0 = (0,0,0,-k)$ of the $P^{(k_0,s,\eta)-}$ representation. We will show that in the $P^{(k_0,s,\eta)-}$ representation, time reflections prevent that SU(2) commutes with $Z_2$. In spite of SU(2) does not commute with $Z_2$, in the $P^{(k_0,s,\eta)-}$ representation, still $\eta$ is a label for the full Poincare group representation. Since the Casimir operators of the full Poincare groups are $P^2$, $W^2$, and $I^2$.

If the vector $|\hat{k}\rangle$ is any vector that transforms according to a representation of the group of translations, then

$$\hat{P}|\hat{k}\rangle = -i\hat{k}|\hat{k}\rangle \tag{12}$$

Hence, in the $P^{(k_0,s,\eta)+}$ representation

$$\hat{P}|\hat{k}_0 s m_s \eta\rangle = -ik|\hat{k}_0 s m_s \eta\rangle \tag{13}$$

While in the $P^{(k_0,s,\eta)-}$ representation

$$\hat{P}|I_t \hat{k}_0 s m_s \eta\rangle = +ik|I_t \hat{k}_0 s m_s \eta\rangle \tag{14}$$

Then, time reflection induces a change on the energy sign, that is

$$I_t P_0 I_t^{-1} = -P_0 \tag{15}$$

That is, time inversions do not commute with $P_0$.

The Pauli-Lubanki four vector components on the rest frame in the $P^{(k_0,s,\eta)+}$ representation is

$$W_q |\hat{k}_0 s m_s \eta\rangle = -k J_q |\hat{k}_0 s m_s \eta\rangle, \text{ and}$$

$$W_t = 0, q = x, y, z \tag{16}$$

Now, in the $P^{(k_0,s,\eta)-}$ representation, on the orbit of the vector $I_t \hat{k}_0 = (0,0,0,-k)$ yields

$$W_q \left| I_t \hat{k}_0 s m_s \eta \right\rangle = +k J_q \left| I_t \hat{k}_0 s m_s \eta \right\rangle, \text{and}$$

(17)

$$W_t = 0, q = x, y, z$$

So that, time inversions do not commute with the Pauli-Lubansky four-vector in the larger space composed by $P^{(k_0,s,\eta)+}$ and $P^{(k_0,s,\eta)-}$.

$$I_t W_q I_t^{-1} = -W_q,$$

(18)

Therefore, from equation (1616) one obtains

$$I_t J_q I_t^{-1} = -J_q$$

(19)

That is, time reversal induces an inversion on the direction of a rotation and changes the sign of the rest energy (rest mass) of the particle. Then, by using equation (1919) one gets

$$J_z I_t \left| \hat{k}_0 s m_s \eta \right\rangle = -I_t J_z \left| \hat{k}_0 s m_s \eta \right\rangle = I_t \left| \hat{k}_0 s m_s \eta \right\rangle (-m_s)$$

(20)

The vector $I_t \left| \hat{k}_0 s m_s \eta \right\rangle$ transforms like $-m_s$, under rotations about the z-axis. Therefore from equation (1919) for a general rotation, we get

$$T(R(\theta))T(I_t) \left| \hat{k}_0 s m_s \eta \right\rangle = T(I_t)T(R(-\theta)) \left| \hat{k}_0 s m_s \eta \right\rangle =$$

(21)

$$T(I_t) \sum_{m'_s} \left| \hat{k}_0 s m'_s \eta \right\rangle D^{(s)-1}_{m'_s m_s}(\theta) = T(I_t) \sum_{m'_s} \left| \hat{k}_0 s m'_s \eta \right\rangle \widetilde{D^{(s)*}_{m'_s m_s}}(\theta)$$

Hence, the vectors $T(I_t) \left| \hat{k}_0 s m_s \eta \right\rangle$ transform like the transpose conjugate complex, $D^{(s)*}_{m'_s m_s}$, of the representation $D^{(s)}_{m'_s m_s}$ of the unitary irreducible representation of SU(2). This *fact explains why it is necessary to conjugate and transpose the wave equation of a particle to describe its antiparticle*. Thus, time reflection induces negative energy states and these states induce charge conjugation. If a mirror reflection is a symmetry transformation, then this reflection must be accompanied by simultaneous space-time inversions, since the intrinsic parity label is generated by space reflection.

Due to the fact that one has to conjugate and transpose, time reversal acquires the properties of an antilinear operator. And since the character of the representations of the three-dimensional rotation group is real, the representations, $D^{(s)*}_{m'_s m_s}$ and $D^{(s)}_{m'_s m_s}$, are equivalent and those representations can be reached one from the other by a change of basis. We apply this transform to the transpose of the transpose of complex conjugated representation. On the new basis the basis vector is given by

$$\left|\hat{k}_0 s m_s \eta\right\rangle_{\substack{new \\ basis}} = \sum_{m_s'} \delta^{m_s'}{}_{-m_s} (-)^{s-m_s'} \left|\hat{k}_0 s m_s' \eta\right\rangle \tag{22}$$

Since the matrix $D^{(s)}$ transforms into the matrix $D^{(s)*}$ according to a set of similarity transformations. As a matter of fact we are mapping a vector over its dual. Then, one has to define

$$\widetilde{D^{(s)*}_{m_s' m_s}} = F^{-1} D^{(s)}_{m_s' m_s} F^1 \tag{23}$$

where the transformation matrix $F$ is given by

$$F = \delta^{m_s'}{}_{-m_s} (-)^{s-m_s} \tag{24}$$

By applying the linear time reversal operator, on the new basis vector, to the general basis vector $\left|k s m_s \eta\right\rangle$ which, is defined by equation (55), one gets

$$T(I_t)\left|\hat{k} s m_s \eta\right\rangle = \sum_{m_s'} T(I_t)\left|\hat{k} s m_s' \eta\right\rangle \delta^{m_s'}{}_{-m_s} (-)^{s-m_s'} = \left|I_t \hat{k} s - m_s \eta\right\rangle (-)^{s+m_s} \tag{25}$$

Under the action of the time reversal operator a particle on the rest frame reverses its energy, and spin.

We define the scalar product according to reference [13] by

$$\left|\psi\right\rangle = \sum_{m_s} \int \psi_{m_s}(\hat{k}) \left|\hat{k} s m_s\right\rangle k_t^{-1} dk \tag{26}$$

Therefore applying the bra-vector of equation (25) to equation (2626) one gets

$$\psi'_{m_s}(\hat{k}) = (-1)^{s-m_s} \psi^*_{-m_s}(I_t \hat{k}) \tag{27}$$

Here, the conjugation is a consequence of the bra-vector. Applying the space reversal operator to the general basis vector $\left|k s m_s \eta\right\rangle$, one obtains

$$T(I_s)\left|\hat{k} s m_s \eta\right\rangle = T(W_{-k})T(I_s)\left|\hat{k}_0 s m \eta_s\right\rangle = T(W_{-k})\left|\hat{k}_0 s m_s \eta\right\rangle \eta = \left|I_s \hat{k} s m \eta_s\right\rangle \eta \tag{28}$$

The momentum is reversed and the particle acquires a definite parity. In order to generate the basis vectors of the representation for the Poincare group with simultaneous space-time reflections, we apply the time reversal operator to equation (99)

$$T(I)|\hat{k}sm_s\eta\rangle = T(I_t)T(I_s)T(W_k)|\hat{k}_0sm_s\eta\rangle = T(I_t)T(W_{-k})|\hat{k}_0sm_s\eta\rangle\eta$$
$$= T(I_t)|I_s\hat{k}sm_s\eta\rangle\eta = |I\hat{k}s-m_s\eta\rangle\eta(-)^{s+m_s} \qquad (29)$$

Therefore, in this formulation the simultaneous action of space-time inversions on a general basis vector (particle) of the representation reverses the energy, momentum, and spin (antiparticle). The space inversion furnishes a definite parity, η, to the elementary particle described by the unitary irreducible representation. The elementary particle violates the causality principle because it is represented in the negative energy sector of the light cone. If it has enough energy it can be observed as a particle moving backwards in space and time (antiparticle). Therefore, the C, T, P, CT, CP, PT cannot conserve separately, but CPT. In my opinion not all the symmetries that would enhance the full Poincare group are known. Then, it is quite possible that exists a violation of CPT for physical phenomena that requires more large symmetries. The full group of diffeomorphisms (space-time reflections included) is a larger symmetry than that of the full Poincare group; therefore CPT could be violated by quantum gravity. By the same token supersymmetry, supergravity, and superstrings could violate CPT.

From equation (29), we get

$$T^2(I)|\hat{k}sm_s\eta\rangle = |\hat{k}sm_s\eta\rangle(-)^{2s} \qquad (30)$$

Hence, $T^2 = 1$ for integer spin, and $T^2 = -1$ for particles of half integer spin.

**4. Particles of Zero Mass**

For massless particles, or the future null and past null regions 4 and 5, the unitary irreducible representations of the Poincare group are labeled by unitary irreducible representations of its little group. The little group of that of Poincare group in these regions is the Euclidean group $E^{(2)}$ in two-dimensions, one will denote its elements by $W_{k_0}$. The $W_{k_0}$'s are the transformations that leave invariant the four-vector $\hat{k}_0$. The orbit of the vector $\hat{k}_0 = (0,0,1,1)$ is the light cone surface. Let $W_k$ be a Lorentz transformation that takes $\hat{k}_0$ into $\hat{k}$, that is $W_k\hat{k}_0 = \hat{k}$; then a general vector can be written by

$$|\hat{k}m\rangle = T(W_k)|\hat{k}_0 m\rangle. \qquad (31)$$

It can be proved that $W_{k'}^{-1}WW_k$ is in the little group, that is $W_{k_0} = W_{k'}^{-1}WW_k$, then applying a general Lorentz transformation $W$ followed by a translation $\hat{u}$ one obtains.

$$T(\hat{u})T(W)|\hat{k}m\rangle = |\hat{k}'m\rangle \exp(ik',\hat{u})\exp(-im\theta) \tag{32}$$

The $E^{(2)}$ representations with continuous spin were ignored in this revision. We took on consideration only the representations labeled by the two-dimensional rotation group, that is $m = 0, \pm\frac{1}{2}, \pm 1..., etc.$

By enlarging the Poincare group to include space-time inversions the little group of the vector $\hat{k}_0 = (0,0,1,1)$ $\widetilde{E}^{(2)} = Z_2 \times E^{(2)}$. Hence, the "full" Poincare group is labeled by η = ±1 and a label of *SO(2)*, because the continuous spin representations of $E^{(2)}$ will not be considered in these regions of the light cone, then $E^{(2)}$ is labeled by *SO(2)*. If one applies space-time inversion over the basis vector

$$|\hat{k}m\rangle = T(W_k)|\hat{k}_0 m\rangle \tag{33}$$

Where $W_k\hat{k}_0 = \hat{k}$, furthermore if one notices that under space inversions

$$(W_q - imp_q)T(I_s)|\hat{k}m\rangle = T(I_s)(W_q + imp_q)|\hat{k}m\rangle = 0 \tag{34}$$

$$(W_t - imp_t)T(I_s)|\hat{k}m\rangle = T(I_s)(W_t + imp_t)|\hat{k}m\rangle = 0 \tag{35}$$

Hence,

$$T(I_s)|\hat{k}m\rangle = |I_s\hat{k} - m\rangle \tag{36}$$

and

$$T(I_s)|\hat{k} - m\rangle = |I_s\hat{k}m\rangle \tag{37}$$

Therefore, these vectors belong to the $P^{(0,m)} + P^{(0,-m)}$ representation, contrary to the representation $P^{(k,m)}$ these representations do not carry spin, but helicity. Therefore, time reversal does not induce a change in sign of the angular momentum *J,* as it does in equation (1919).

Also, if one notices that

$$(W_q + imp_q)T(I_t)T(I_s)|\hat{k}m\rangle = T(I_t)(W_q - imp_q)T(I_s)|\hat{k}m\rangle = 0$$

$$(W_t + imp_t)T(I_t)T(I_s)|\hat{k}m\rangle = T(I_t)(W_t - imp_t)T(I_s)|\hat{k}m\rangle = 0 \tag{38}$$

Hence,

$$T(I_t)T(I_s)|\hat{k}m\rangle = |I_t I_s \hat{k}m\rangle \tag{39}$$

Therefore, under space-time inversions the vector $T(I)|\hat{k}m\rangle$ transforms according to the representation $T^{(Ik)}$. Additionally, it can be proved that such a vector transforms in the same manner under translations. Now, according to equations,

$$(W_q + imp_q)|\hat{k}m\rangle = 0$$

$$(W_t + imp_t)|\hat{k}m\rangle = 0, \tag{40}$$

the $|I\hat{k}m\rangle$ vector transforms according to the $P^{(0,m)}$ representation, but with its energy and momentum reversed.

On the $|\hat{k}_0 0\eta\rangle$ representation of the little group of $\hat{k}_0$ one gets

$$T(I)|\hat{k}0\eta\rangle = T(W_k)T(I_t)T(I_s)|\hat{k}_0 0\eta\rangle = |I\hat{k}_0 0\eta\rangle\eta \tag{41}$$

This equation represents a particle of zero rest mass, and zero helicity, but with a definite parity that changes its momentum and energy under space-time reversal.

We realize that transitions between time-like and space-like states for a "massless" particle are highly probable, since if the four momentum, $p^2 = 0$, is sharply defined, then its four vector position

$$(x_1 - x_2)^2 - (x_1^0 - x_2^0)^2 \sim \frac{\hbar^2}{p^2 \to 0} \qquad (42)$$

is completely uncertain, since its Compton wavelength is undetermined [14]. The massless particle could exist in any place of the light cone. The principle of causality is strongly violated. If one observes a given amount of photons with $p^2 = 0$ is because a big amount of other photons populate the space-like, and the time-like regions. The action, of these dark photons, on the universe should be measurable.

## 5. Imaginary Mass Particles

In the space-like, region 1, the eigenvalue of the square of the four-momentum is the negative of the square of the rest mass of the particle. Therefore, the mass would be imaginary number. No physical interpretation of elementary particles could be associated to this region; however, the square of the four-momentum could have a physical meaning. It turn out that the square of the four-momentum is not equal to the square of the mass for internal lines in a Feynman diagram. It is conjectured in this article that a possible interpretation could be given to the unitary irreducible representations of the "full" Poincare group. Virtual particles would be good candidates to achieve this goal.

The little group of the four-vector $\hat{k}_0 = (0,0,k,0)$ is the SL(2,r) which contains spin representations. These representations where classified independently by Bargmann, Naimark and Gelfand, and Harish-Chandra[13].

It can be shown that a general basis vector is given by

$$\left| \hat{k}m \right\rangle = T(W_k) \left| \hat{k}_0 m \right\rangle = T(R_z(\theta))T(W_{k_1})T(W_{k_3}) \left| \hat{k}_0 m \right\rangle \qquad (43)$$

If we apply the space reversal operator to the four basis vector and consider as the little group the group of reflections and SL(2,R), $Z_2 X SL(2,R)$, one obtains

$$T(I_s) \left| \hat{k}sm_s \eta \right\rangle = \sum_{m_s'} T(I_s) \left| \hat{k}sm_s' \eta_s \right\rangle \delta^{m_s'}_{-m_s} (-)^{s-m_s'} = \left| I_s \hat{k} s - m_s \eta \right\rangle (-)^{s+m_s}, \qquad (44)$$

since the relation

$$I_s J_z I_s^{-1} = -J_z, \qquad (45)$$

is obtained when one chooses the point $\hat{k}_0 = (0,0,-z,0)$ to construct the space-like representations. Applying time reversal to the general basis vector, one finds that the four-vector $\hat{k}$ reverses its spatial components, by following a similar procedure than that of section 3.

$$T(I_t)\left|\hat{k}m\eta\right\rangle = T(W_{-k})T(I_t)\left|\hat{k}_0 m\eta\right\rangle = \left|I_s\hat{k}m\eta\right\rangle\eta \tag{46}$$

Hence, the action of space-time inversion yields

$$T(I)\left|\hat{k}sm_s\eta\right\rangle = T(I_t)T(I_s)T(W_k)\left|\hat{k}_0 sm_s\eta\right\rangle = T(I_s)T(W_{-k})\left|\hat{k}_0 sm_s\eta\right\rangle\eta$$
$$= T(I_s)\left|I_s\hat{k}sm_s\eta\right\rangle\eta = \left|\hat{k}s - m_s\eta\right\rangle\eta(-)^{s+m_s} \tag{47}$$

The vectors of the discrete series $D^+$ transform into the $D^-$ discrete series, and vice versa.

## 6. Vacuum Representations

In region six the little group of the four-vector $\hat{k}_0 = (0,0,0,0)$ is the Lorentz group. Because the Lorentz group is non-compact, its unitary irreducible representations are infinite dimensional. Therefore, the unitary irreducible representations of the Poincare group in this region are the same than that of the Lorentz group. They are labeled by the eigenvalues of the Casimir of the algebra SU(2)XSU(2), that is, the angular momentum on three dimensions u, v. The group SU(2)XSU(2) is locally isomorphic to the Lorentz group, therefore their Lie algebras are equivalents. There are two classes of these representations, the principal series ($v = -iw$, $w$ real, $j_0 = 0, 1/2, 1, …$) and the complementary series ($-1 \leq v \leq 1$; $j_0 = 0$). Because combined space-time transformations commute with the generators of the Lorentz group the states of this representation have a definite parity. Furthermore, vacuum oscillations make these states to undergo transitions to positive and negative energy states as well as imaginary mass (virtual states).

## Conclusion

This paper proves that the interpretation of tachyons as antiparticles, set on a realistic basis Sudarshan's conjecture of the existence of particles that move faster than the speed of light, that is, superluminal particles.

## Acknowledgments

I would like to thank George Sudarshan for discussing with me the incompatibility of Bose-Einstein and Fermi-Dirac statistics to explain the creation of antiparticles under the same theoretical scheme. I am in debted with Yuval Ne'eman for help me to understand the importance of group theory in physics. The last acknowledgment, but not the least goes to Roger Pecina for his worthy commentaries to the manuscript.